\documentclass[aps,prl,floatfix,showpacs,superscriptaddress,notitlepage,twocolumn,
amsmath,amssymb]{revtex4}

\usepackage{graphicx}
\usepackage{bm}
\usepackage{braket}
\usepackage{times}
\expandafter\let\csname equation*\endcsname\relax
\expandafter\let\csname endequation*\endcsname\relax
\usepackage{amsmath}
\usepackage{amssymb}
\usepackage{amstext}

\newcommand{\ER}{E_R}
\newcommand{\Fig}[1]{Fig.~\ref{#1}}

\usepackage{color}
\newcommand{\cites}[1]{{\color{blue}\cite{#1}\color{black}}}
\newcommand{\X}{{(2n-1)a_s\,\Delta J}}
\newcommand{\Xone}{{a_s\,\Delta J}}

\begin{document}

\title{Observation of Density-Induced Tunneling}

\newcommand{\ILP}{\affiliation{Institut f\"ur Laser-Physik, Universit\"at Hamburg, Luruper Chaussee 149, 22761 Hamburg, Germany}}
\newcommand{\UI}{\affiliation{Institut f\"ur Experimentalphysik und Zentrum f\"ur Quantenphysik, Universit\"at Innsbruck, 6020 Innsbruck, Austria}}

\author{Ole J\"urgensen}\ILP
\author{Florian Meinert}\UI
\author{Manfred J. Mark}\UI
\author{Hanns-Christoph N\"agerl}\UI
\author{Dirk-S\"oren L\"uhmann}\ILP


\begin{abstract}
We study the dynamics of bosonic atoms in a tilted one-dimensional optical lattice and report on the first direct observation of density-induced tunneling. We show that the interaction affects the time evolution of the doublon oscillation via density-induced tunneling and pinpoint its density- and interaction-dependence. The experimental data for different lattice depths are in good agreement with our theoretical model.
Furthermore, resonances caused by second-order tunneling processes are studied, where the density-induced tunneling breaks the symmetric behavior for attractive and repulsive interactions predicted by the Hubbard model.
\end{abstract}


\maketitle
The Hubbard model is the primary description for strongly correlated electrons in solids. It takes into account the interaction between the particles at a lattice site and the tunneling between the sites, whereas other interaction processes are neglected. It was pointed out that these additional interactions may have crucial influence in strongly correlated materials such as superconductors or ferromagnets \cites{Hirsch1989, Strack1993, Hirsch1994, Amadon1996}. Of particular importance is the so-called bond-charge interaction that represents a \textit{density-induced tunneling} of an electron. In solids, this interaction-driven process cannot be studied systematically due to the lack of direct control over the electron density and the interaction strength. Furthermore, the complexity of the investigated materials hinders a direct observation of this interaction effect. Hence, the role of interaction-induced tunneling has remained an open question in condensed matter physics.

Ultracold atoms in optical lattices allow the realization of extremely pure lattice systems without defects and phononic excitations. Furthermore, the unique control of both the lattice potential and the interaction strength permits a systematic study of static and dynamic properties. In optical lattices, density-induced tunneling \cites{Mazzarella2006, Mering2011, Luhmann2012, Jurgensen2012,Dutta2014} is even more pronounced due to the characteristic shape of the Wannier functions \cites{Luhmann2012}. Several indications for density-induced tunneling have been found: It has a strong influence on the superfluid to Mott-insulator transition in bosonic \cites{Luhmann2012, Pilati2012, Mark2011} and multicomponent systems such as Bose-Fermi mixtures of atoms \cites{Ospelkaus2006, Gunter2006, Best2009, Mering2011, Jurgensen2012, Luhmann2008}. As a tunneling process, it also modifies the effective band structure, which has also been observed in a Bose-Fermi mixture \cites{Heinze2011}. A direct observation of density-induced tunneling processes was hindered mainly by the fact that the Mott insulator transition depends only on the ratio of on-site interaction and \textit{total} tunneling and by the averaging over different on-site occupancies in experimental systems.

\begin{figure}[b]
\includegraphics[width=\linewidth]{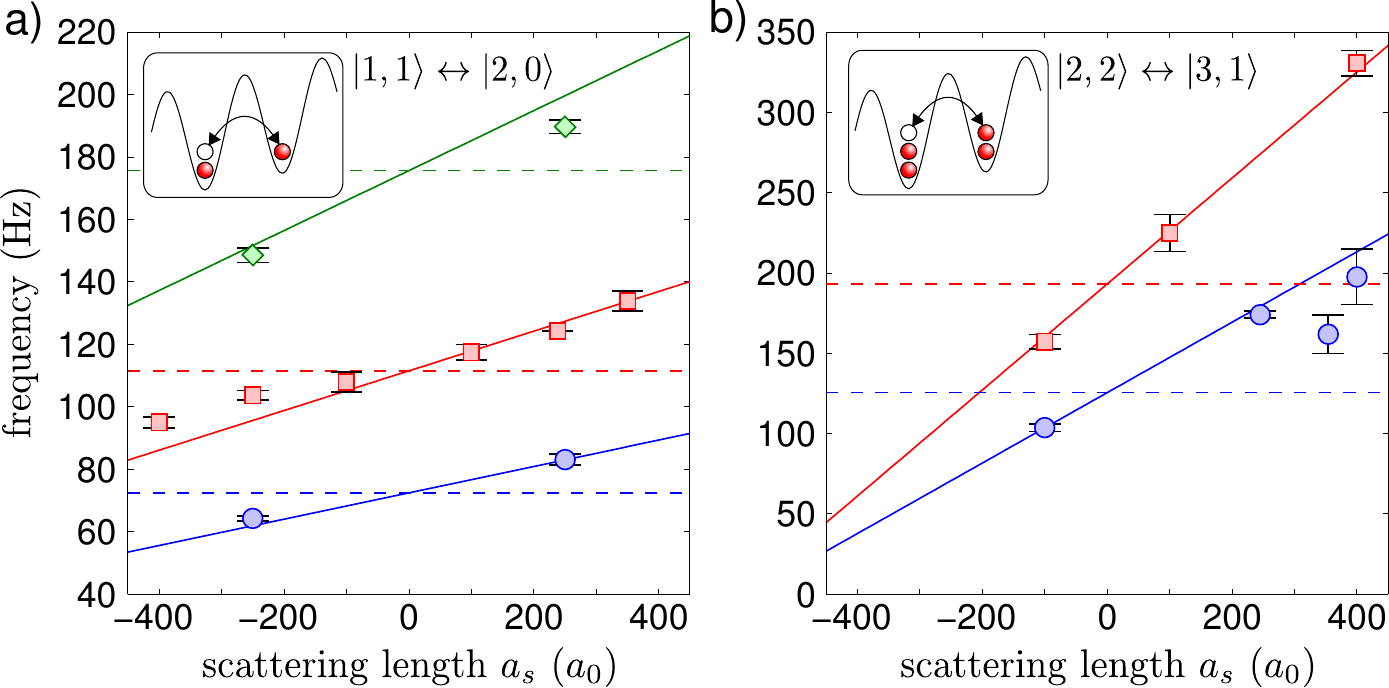}
\caption{ Oscillation frequency $f_0$ of the doublon number in a tilted lattice as a function of $a_s$ for $V_z=8\ER$ (diamonds), $10\ER$ (squares), and $12\ER$ (circles) with initial on-site occupancy (a) $n=1$ and (b) $n=2$ (see insets). The dashed lines show the prediction of the standard Hubbard model. The solid lines depict the interaction dependence due to density-induced tunneling (see Eq.~\eqref{eq_overallTunneling}). Theoretical predictions correspond to a $4\%$ lower lattice depth \cites{SuppMat}. The experimental data of (a) is taken from Ref.~\cites{Meinert2013}.
}
\label{fig_ExpTh}
\end{figure}

Here, we report on the direct observation of density-induced tunneling with ultracold atoms in a tilted one dimensional optical lattice. We study the dynamics of a 1D Mott insulator after quenching the tilt energy $E$ between neighboring sites into resonance with the on-site interaction energy $U$. We show that the resulting resonant particle oscillation between neighboring sites (see inset of \Fig{fig_ExpTh}) is driven by interaction-induced tunneling on top of conventional tunneling. The experimental control over the on-site occupancy and the interaction strength via a Feshbach resonance \cites{Mark2011} allows us to isolate the effect of interaction-induced tunneling and to study it systematically.

The observed oscillation frequency $f_0$ is expected to be directly proportional to the tunneling matrix element $J$ and is plotted in \Fig{fig_ExpTh} as a function of the atomic scattering length $a_s$. The plot shows that the tunneling rate  is modified by the interaction strength and the density of the sample. The Hubbard model predicts a \textit{constant} value for $J$ (dashed lines in \Fig{fig_ExpTh}). In contrast, density-induced tunneling $\Delta J$ changes linearly with $a_s$ and the on-site occupancy $n$, contributing to the total tunneling energy via
\begin{equation} \label{eq_overallTunneling}
J_\text{tot} = J + \X.
\end{equation}
The measured oscillation frequency agrees very well with this modified tunneling rate (solid lines in \Fig{fig_ExpTh}). An increase of the on-site occupation from $n=1$ to $n=2$ increases the slope of the experimental data in accordance with Eq.~\ref{eq_overallTunneling}. Here, the  amplitude of density-induced tunneling can even be as strong as the one of conventional tunneling.


The experiments are performed as reported in detail in Ref. \cites{Meinert2013}. Starting from a Cs Bose-Einstein condensate, we prepare a bosonic Mott insulator in a 3D cubic optical lattice with  a lattice depth $V_q = 20\,E_R$ ($q=x,y,z$), where $E_R= h \times 1.325\,$kHz denotes the photon recoil energy, and $h$ is Planck's constant \cites{SuppMat}. Adjusting initial density, interactions, and external confinement during lattice loading allows us to prepare either a clean one-atom- or a two-atom-per-site Mott shell \cites{SuppMat}. We set $a_s$ to the desired value ($-400\,a_0 \leq a_s \leq +400\,a_0$) by means of a Feshbach resonance. Tunneling dynamics along 1D chains is initiated by first setting the tilt $E$ along the vertical $z$-direction via a magnetic force $|\nabla B|$ and then quickly lowering $V_z$ along the direction of the tilt. We measure the number of doubly occupied sites (doublons) after a variable evolution time $t_h$ through Feshbach molecule formation and detection \cites{Meinert2013}. On resonance ($E\approx U$) the doublon number exhibits large-amplitude oscillations. For $n=1$ the oscillation frequency $f_0$ is deduced from a damped sinusoidal fit to the data \cites{Meinert2013}. In \Fig{fig_ExpTh}(a) we give $f_0$ as a function of $a_s$ for three different $V_z$ from data sets taken for Ref. \cites{Meinert2013}. Time traces for $n=2$ need a more refined spectral analysis (see below). \Fig{fig_ExpTh}(b) plots the (mean) frequency deduced from measurements with $n=2$ as a function of $a_s$ for $V_z=10\, E_R$ and $V_z=12\, E_R$. The observed frequencies clearly depend on both the interaction strength $a_s$ and the on-site occupancy $n$ in the lattice. The Hubbard model (dashed lines) predicts constant $J$, which solely depends on the lattice depth $V_z$ \cites{Jaksch1998}, and cannot reproduce this behavior.

\begin{figure}
 \includegraphics[width=\linewidth]{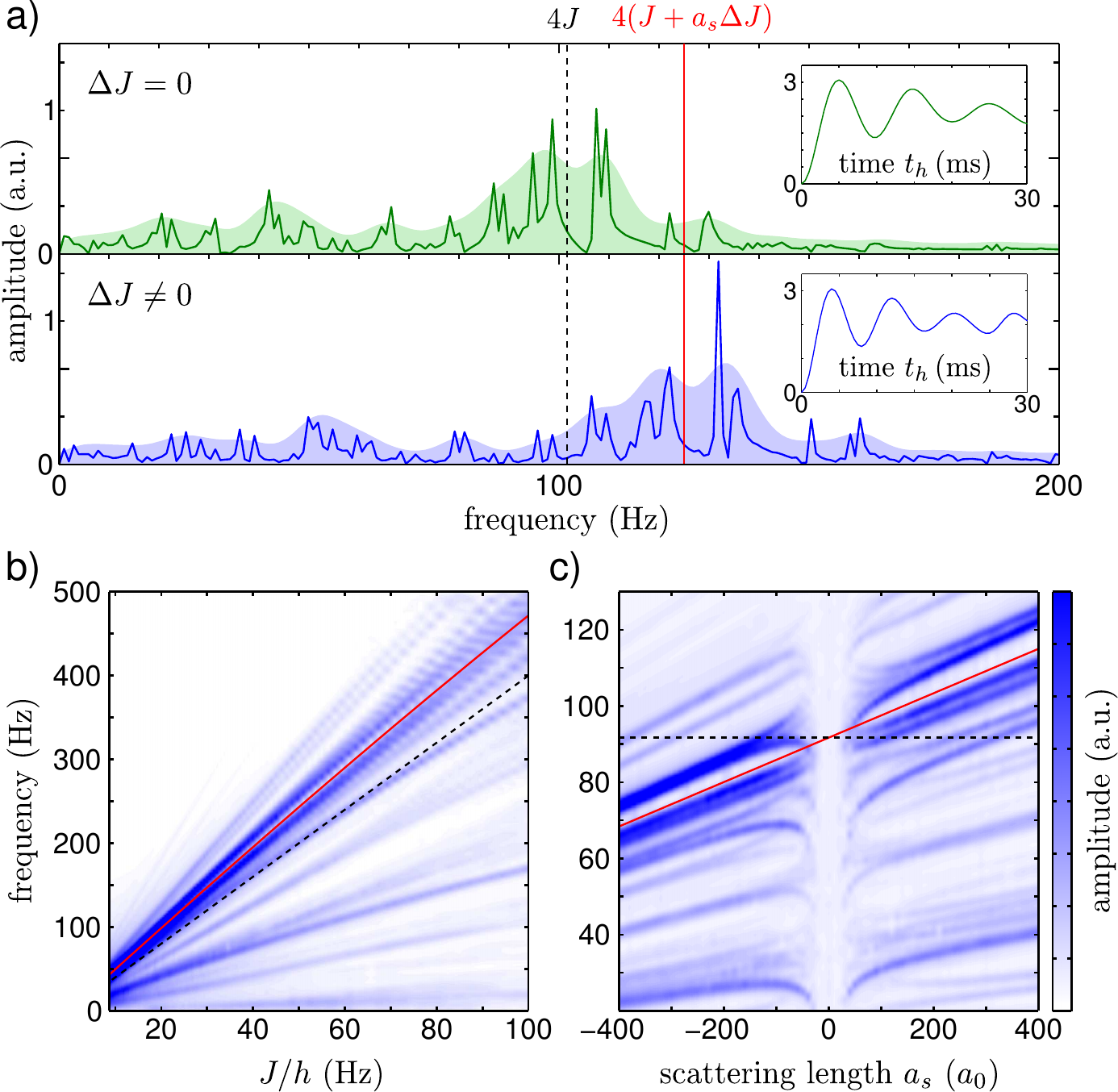}
\caption{
(a) Fourier spectrum of the simulated doublon number dynamics for $E=U$ without (upper panel) and with (lower panel) inclusion of density-induced tunneling. Here, $n=1$, $V_z=10 E_R$, and $a_s = 400\, a_0$. The dashed (solid) line shows the predominant frequency component predicted without (with) density induced tunneling. The shaded areas incorporate a Gaussian broadening with a width $w =4\,\text{Hz}$. The insets show the time trace for the first $30\, \text{ms}$. (b) Mode spectrum as a function of the single-particle tunneling rate $J/h$ for $a_s=400 a_0$ ($w =4\,\text{Hz}$). (c) Mode spectrum as a function of $a_s$ at fixed $V_z=10 E_R$ ($w =1\,\text{Hz}$). In (b) and  (c) the dashed line shows $4 J/h$, while the solid line shows $4 J_{\rm{tot}}/h$, including density-induced tunneling.
}
\label{fig_spectra}
\end{figure}

\begin{figure*}
\centering
 \includegraphics[width=0.95\linewidth]{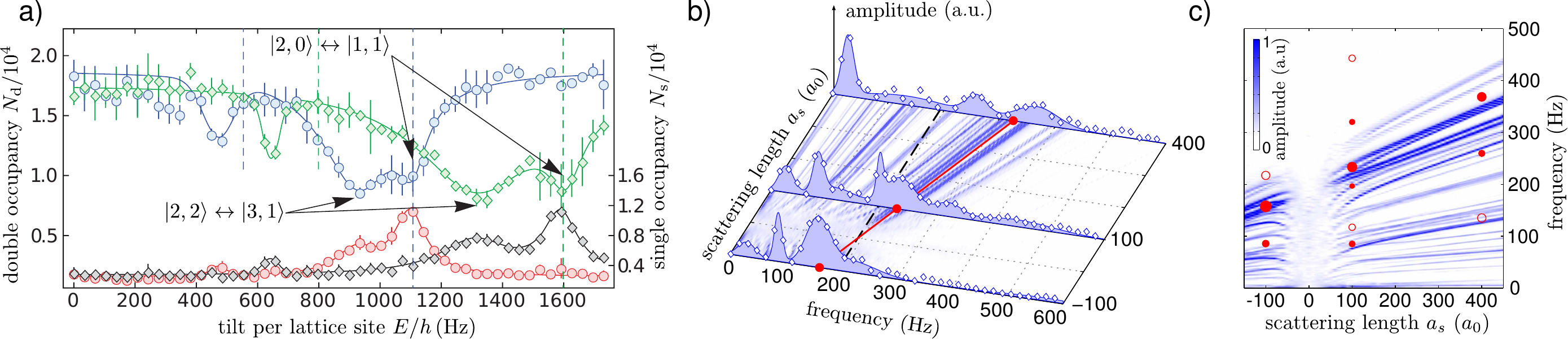}
\caption{Doublon dynamics for on-site occupancy $n=2$. (a) Number of doubly (upper lines) and singly (lower lines) occupied sites as a function of $E$ after $t_h=50\,\mathrm{ms}$ for $a_s=245(5)\,a_0$ (circles) and $a_s=354(5)\, a_0$  (diamonds) at $V_z=12 \ER$, giving $U=h  \times 1107(20) \, \rm{Hz}$ and $U=h \times 1598(20) \, \rm{Hz}$, respectively. The dashed lines indicate the calculated $U$ and $U/2$. (b) Measured Fourier spectra extracted from time traces of the doublon number at $V_z=10 \ER$ for different $a_s$. The profiles result from multiple Gaussians fits to the data \cites{SuppMat}. The shaded area projected into the axis plane depicts the simulated Fourier spectra shown in (c). The lines show the expected frequencies for the standard (dashed) and extended (solid) Hubbard model (Eq.~\eqref{eq_overallTunneling}). (c) Simulated Fourier spectra as a function of $a_s$ incorporating a Gaussian broadening with a width $w=1\, \mathrm{Hz}$. The markers denote peak positions deduced from the experimental data shown in (b). The marker size scales linearly with the area under the respective fits. Open circles indicate broad Gaussian fits with a large ratio of width and amplitude $w/A > 0.2$.}
\label{fig_triplonDynamics}
\end{figure*}

For the theoretical description of the experiment we make use of the generalized Hubbard Hamiltonian for the one-dimensional lattice including density-induced tunneling, which is given by
\begin{eqnarray}\label{eq_Hubbard}
 \hat H&\!\! =\!\!& -J \sum_i \hat b_i^\dagger \hat b_{i+1} +c.c. + \frac{U}{2} \sum_i \hat n_i (\hat n_i -1) + E \sum_i \hat n_i i \nonumber\\
	& &-  a_s\,\Delta J \ \sum_i \hat b_i^\dagger (\hat n_i + \hat n_{i+1}) \hat b_{i+1} + c.c.
\end{eqnarray}
with the tunneling matrix element $J$, the on-site interaction $U$, a  tilt $E$ per lattice site, bosonic annihilation (creation) operators $\hat b_i^{(\dagger)}$ on site $i$, and $\hat n_i=\hat b_i^\dagger \hat b_i$.
The first line is the standard Bose-Hubbard model, whereas the second line represents the density-induced tunneling operator. It originates from the two-body interaction operator and represents the dominant off-site contribution for neutral atoms in optical lattices \cites{Luhmann2012, Jurgensen2012, Dutta2014}. Its amplitude
\begin{equation}
 \Delta J = - \frac{4 \pi \hbar^2 }{m}  \int d^3 r\ w^* (\mathbf{r-d}) w^* (\mathbf{r}) w^2 (\mathbf{r}),
\end{equation}
is determined by the Wannier functions $w(\mathbf{r})$ of the lowest band of the lattice with the lattice spacing $d$ using a $\delta$-shaped interaction potential. This tunneling operator is explicitly occupation-dependent due to the factor $(\hat n_i + \hat n_{i+1})$. Assuming that the time-evolution on neighboring sites is predominantly given by a constant total occupation $n_i+n_{i+1}=2n$, we can define an effective total tunneling operator as
\begin{equation} \label{eq_overallTunneling_operator}
\hat J_\mathrm{tot} = - J_\text{tot} \sum_i \hat b_i^\dagger \hat b_{i+1} +c.c.
\end{equation}
using Eq.~\eqref{eq_overallTunneling}, which allows us to retrieve the standard Bose-Hubbard model with a modified tunneling rate $J_{\rm{tot}}$.

It is a priori not clear that conventional and density-induced tunneling can be combined to one total hopping process. To verify this simplification, we perform exact numerical simulations of the time evolution of the initial state by diagonalizing the generalized Hubbard model \eqref{eq_Hubbard} for a finite  lattice with $N=8$ sites.  Using the exact solution, we are not restricted to short time traces, allowing us to resolve the full Fourier spectrum. We first discuss the case of an initial on-site occupation $n=1$. As an example, the insets in \Fig{fig_spectra}(a) show time traces of the number of doublons ($n_i=2$) for $V_z=10\ER$ and $a_s=400 a_0$ at the resonance $E=U$. Here, the number of triply occupied sites (triplons) is negligible. The Fourier spectrum (\Fig{fig_spectra}(a)) contains a broad range of frequencies that are peaked around $f_0=\nu J_\mathrm{tot} /h$ (bottom) or around $f_0= \nu J/h$ (top) for the standard Hubbard model ($\Delta J = 0$) with the prefactor $\nu\approx 4$, as indicated by the vertical lines. By means of time-dependent DMRG of up to 40 sites, it has been shown in Ref.~\cites{Meinert2013} that the system size does not significantly affect the characteristic oscillation frequency $f_0$ for Mott chains beyond 3 sites but causes increased many-body damping with increasing system size.

In \Fig{fig_spectra}(b), the frequency spectrum is plotted against the bare tunneling rate $J/h$ for the generalized Hubbard model ($\Delta J \neq 0$). The centroid of the two strongest modes matches with the total tunneling rate $4 J_\mathrm{tot} /h$, while the dashed line corresponds to the standard Hubbard model with $4 J /h$. Plotted as a function of the scattering length $a_s$ (\Fig{fig_spectra}(c)), the difference between standard and generalized Hubbard model becomes even more obvious. As the absolute value of the on-site interaction is compensated by the resonance condition $E=U$, the doublon dynamics is independent of $a_s$ within the standard Hubbard model (dashed line). In contrast, the interaction-dependence of density-induced tunneling imprints a linear dependence on the observed frequency modes $f_0\propto J+a_s\,\Delta J$.

Although the theoretical spectrum of the time evolution contains several distinct features, we can conclude that the main frequency can be attributed to an oscillation with $4 J_\mathrm{tot} /h$. Fitting the experimental time traces for $n=1$ with a damped oscillation \cites{Meinert2013} allows us to extract this central frequency plotted in \Fig{fig_ExpTh}(a). The experimental data points pinpoint the dependence on the interaction strength as discussed above and agree well with the generalized model including density-induced tunneling. Note that the interaction-induced admixture of higher-bands will lead to a slightly modified rate for the total tunneling \cites{Luhmann2012,Jurgensen2012,Bissbort2012,Dutta2014}.


Let us now turn to the direct verification of the density dependence of the interaction-driven tunneling process by preparing an initial state with on-site occupancy $n=2$. Due to $J_\text{tot}=\X$, the impact of the density-induced tunneling is expected to increase by a factor of three. In \Fig{fig_triplonDynamics}(a) we plot the number of doubly and singly occupied sites as a function of $E$ measured after $t_h = 50\,\rm{ms}$ for two different values of $U$. Close to the expected resonance position at $E=U$, we observe two minima in the doublon number (\Fig{fig_triplonDynamics}(a)) that can be attributed to the processes $|2,2\rangle \leftrightarrow |3,1\rangle$ and $|1,1\rangle \leftrightarrow |2,0\rangle$. The latter arises from tunneling at residual defects (empty sites) in the $n=2$ shell \cites{SuppMat}. The splitting of the resonance arises from corrections to the on-site energy $U$ due to multi-orbital effects \cites{Buchler2010, Will2010, Mark2011, Mark2012, Luhmann2012} causing an intrinsically occupation dependent on-site energy  $U_n$. While for the defect process $|1,1\rangle \leftrightarrow |2,0\rangle$ the resonance is at $E_{11}=U_2$, the process $|2,2\rangle \leftrightarrow |3,1\rangle$ is resonant at $E_{22}=3U_3-2U_2 < U_2$. For the measurement of the time traces we determine $E_{22}$ for a fixed lattice depth and scattering length. Since the two resonances are not fully separated, we expect (off-resonant) contributions from defect sites to contribute with up-shifted frequencies \cites{Meinert2013}.

In \Fig{fig_triplonDynamics}(b) Fourier spectra extracted from the experimental time traces \cites{SuppMat} are shown for three different values of $a_s$ in the range $-100\, a_0 \leq a_s \leq +400\, a_0$. We obtain the main frequency modes using Gaussian fits (blue areas). While the lowest observable frequency is caused by decoherence and particle loss, we can identify the dominant mode at the expected position $f_0= \nu (J+3 a_s\,\Delta J)$ (red solid line), with $\nu= 4 \sqrt{3}$ for the $|2,2\rangle \leftrightarrow |3,1\rangle$ process.  For the two positive values of $a_s$, where the frequency range leads to better resolved peaks, a splitting of this mode can be observed. This splitting is in general also visible in the theoretical spectrum in \Fig{fig_triplonDynamics}(c), where the circles denote the experimentally extracted modes. Moreover, we can identify a mode with lower frequencies that could probably be assigned to the mode around $\nu\approx 3 \sqrt{3} / 2$ in the numerical spectrum. However, defect sites will effectively lead to decoupled chains with different lengths surrounded by unoccupied sites affecting the prefactor $\nu$.

The extracted dominant frequencies are plotted for $V_z=10\ER$ and $V_z=12\ER$ \cites{SuppMat} in \Fig{fig_ExpTh}(b), where we use the centroid for split resonances. For both lattice depths, we see a good agreement with the theoretical expectation $f_0= 4 \sqrt{3} (J+3 a_s\,\Delta J)$. In combination with results for $n=1$, this serves as direct confirmation of the density dependence of the tunneling.


\begin{figure}
\centering
 \includegraphics[width=\linewidth]{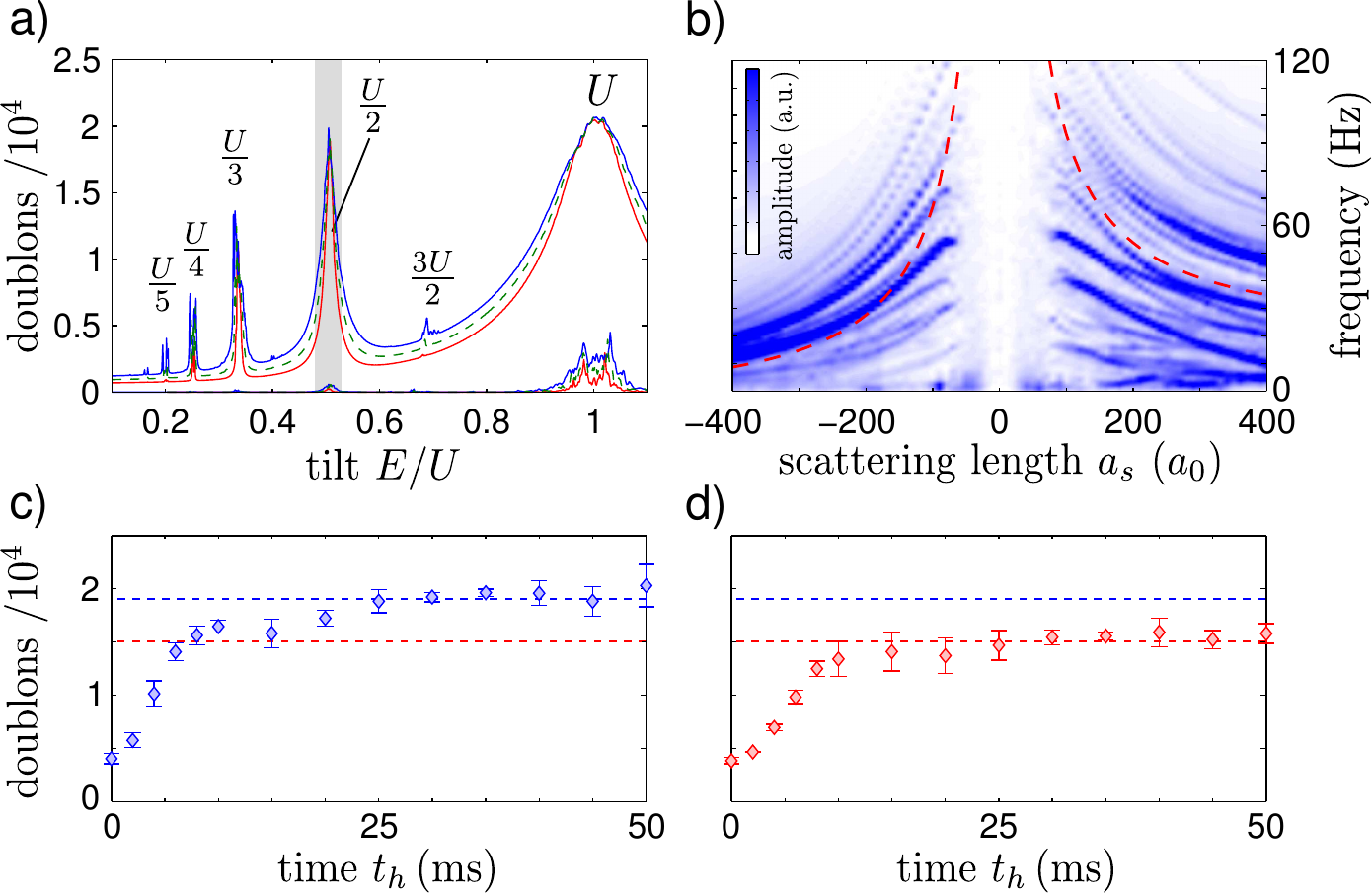}
\caption{(a) Time-averaged number of the sum of doublons and triplons (upper lines) and triplons only (lower lines) as a function of $E$ for $V_z=8\ER$ at $a_s=250 a_0$ (blue) and $a_s=-250 a_0$ (red). The solid lines show numerical simulations including density-induced tunneling, while for the dashed lines $\Delta J = 0$. The gray area indicates sampling over $E$ in the experiment with an estimated width of $\approx 50\,$Hz. (b) Numerical Fourier spectrum of the doublon dynamics as a function of $a_s$ revealing the breaking of symmetry between attractive and repulsive scattering for $E=U/2$. The dashed line indicates $f_0=\nu (J+a_s \Delta J)(J+2 a_s \Delta J)/U$ with a prefactor $\nu=19$ in accordance with \cites{Meinert2013b}. Experimental time traces at the $U/2$ resonance for (c) $a_s=+250 a_0$ and (d) $a_s=-250 a_0$ at $V_z=8\ER$ showing an asymmetric behavior between repulsive and attractive interaction caused by density-induced tunneling. The dashed lines indicate the steady-state doublon number. \label{fig_higherOrder}}
\end{figure}

As has been demonstrated recently \cites{Meinert2013b}, the dynamics in tilted lattices also allows the study of higher-order tunneling processes. In \Fig{fig_higherOrder}(a), the numerically determined doublon number, averaged over the time evolution, is plotted as a function of $E$, depicting several distinct resonances at fractional values of $U$ in accordance with the experimental observation in Ref.~\cites{Meinert2013b}. The resonance at $E=U/2$ is caused by second-order tunneling processes via an intermediate site. In this case, the hopping to next-nearest neighbor sites restores the resonant tunneling condition ($2E=U$), whereas direct nearest-neighbor tunneling is suppressed. Higher-order resonances at $E=U/n$ are caused by long-range tunneling processes proportional to $J (J/U)^{n-1}$. Density-induced tunneling causes a broadening of the resonances for repulsive interactions (blue line) and a narrowing for attractive interactions (red line).

The numerical Fourier spectrum for the resonant second-order tunneling dynamics is plotted in \Fig{fig_higherOrder}(b). A clear evidence for the impact of density-induced tunneling is the breaking of the symmetry between attractive and repulsive interaction, which holds for the Hubbard model, i.e., $\Delta J=0$ \cites{SuppMat}. This frequency shift can also be observed in the experimental time traces in \Fig{fig_higherOrder}(c) and (d), where the faster initial increase for repulsive interactions indicates a higher frequency. In addition, we find a decrease in the average doublon number for attractive interactions, which we attribute to the reduced width of the second-order tunneling resonance (see \Fig{fig_higherOrder}(a)), as the variation of $E$ across the sample due to the small residual harmonic confinement and a variation in $a_s$ due to the magnetic field gradient (gray area) is comparable with the resonance width.


We have presented the first direct measurement of density-induced tunneling of ultracold atoms in optical lattices. We observe resonant doublon dynamics when compensating the interaction energy $U$ by an applied tilt. The measured frequency exhibits a linear dependence on the on-site occupancy and on the scattering length. Our numerical simulations show that an extended Hubbard model incorporating  the density-induced tunneling accurately describes the experiment. For approximately constant densities both tunneling processes can be described with a single effective amplitude $J + \X$ that can differ strongly from the conventional tunneling $J$. Furthermore we have studied second-order tunneling processes and observe an asymmetry between repulsive and attractive interactions caused by density-induced tunneling. This underlines its importance for exchange interactions that are, e.g., responsible for antiferromagnetic properties in solids \cites{Anderson1950}. Our results grants future perspectives for detailed investigations of complex interaction effects caused e.g. by higher orbitals and off-site interactions \cites{Luhmann2012,Bissbort2012}.

We are indebted to R. Grimm for generous support, and thank A. Daley for fruitful discussions. We gratefully acknowledge funding by the Deutsche Forschungsgemeinschaft (grants SFB 925 and GRK 1355) and the European Research Council (ERC) under Project No. 278417.

\onecolumngrid
\vspace{3cm}
\twocolumngrid

\section{Supplementary Material: Observation of Density-Induced Tunneling}
\subsection{Lattice depth calibration and error bars}

The lattice depth $V_q$ is calibrated via Kapitza-Dirac diffraction. The statistical error for $V_q$ is 1\%, though the systematic error can reach up to 5\%.

The scattering length $a_s$ is calculated via its dependence on the magnetic field \cites{Mark2011Sup} with an estimated uncertainty of $\pm 5 a_0$ arising from systematics in the magnetic field calibration and conversion accuracy. Additionally, the magnetic field gradient leads to a variation of less then $\pm 3 a_0$ across the sample.

\subsection{Experimental preparation of the $n=1$ and $n=2$ Mott shell}

Adjusting the chemical potential of the harmonically trapped Bose-Einstein condensate via initial density, interaction strength, and external confinement during the lattice loading allows us to control the relative occupation of the Mott insulating shells with occupation $n=1$ and $n=2$. For the experimental data shown in \Fig{fig_ExpTh}(a) and \Fig{fig_higherOrder} of the main article, we prepare a one-atom-per-site Mott insulator with less than $4\%$ of residual double occupancy. For the data plotted in \Fig{fig_ExpTh}(b) and \Fig{fig_triplonDynamics} of the main article we prepare a Mott insulator with a central two-atom-per-site Mott shell containing $\approx 4.4 \times 10^4$ atoms. We remove the surrounding singly-occupied shell by combining microwave rapid adiabatic passage with a resonant light pulse after association of doubly occupied lattice sites to weakly bound Feshbach molecules. Subsequent to the cleaning, we dissociate the molecules again to free atoms. In total the cleaning procedure has an efficiency of $80(4)\%$ \cites{Meinert2013Sup}.\\

The finite cleaning efficiency results in defects in the $n=2$ shell that are predominantly empty sites or remaining Feshbach molecules, with an estimated defect density of $\approx 20\%$. In combination with tunneling at the chain boundaries this explains the relatively strong resonance corresponding to the process $|1,1\rangle \leftrightarrow |2,0\rangle$ observed in \Fig{fig_triplonDynamics}(a) of the main article.

\subsection{Mode spectrum of the standard Hubbard model}

\begin{figure}[t]
\centering
 \includegraphics[width=\linewidth]{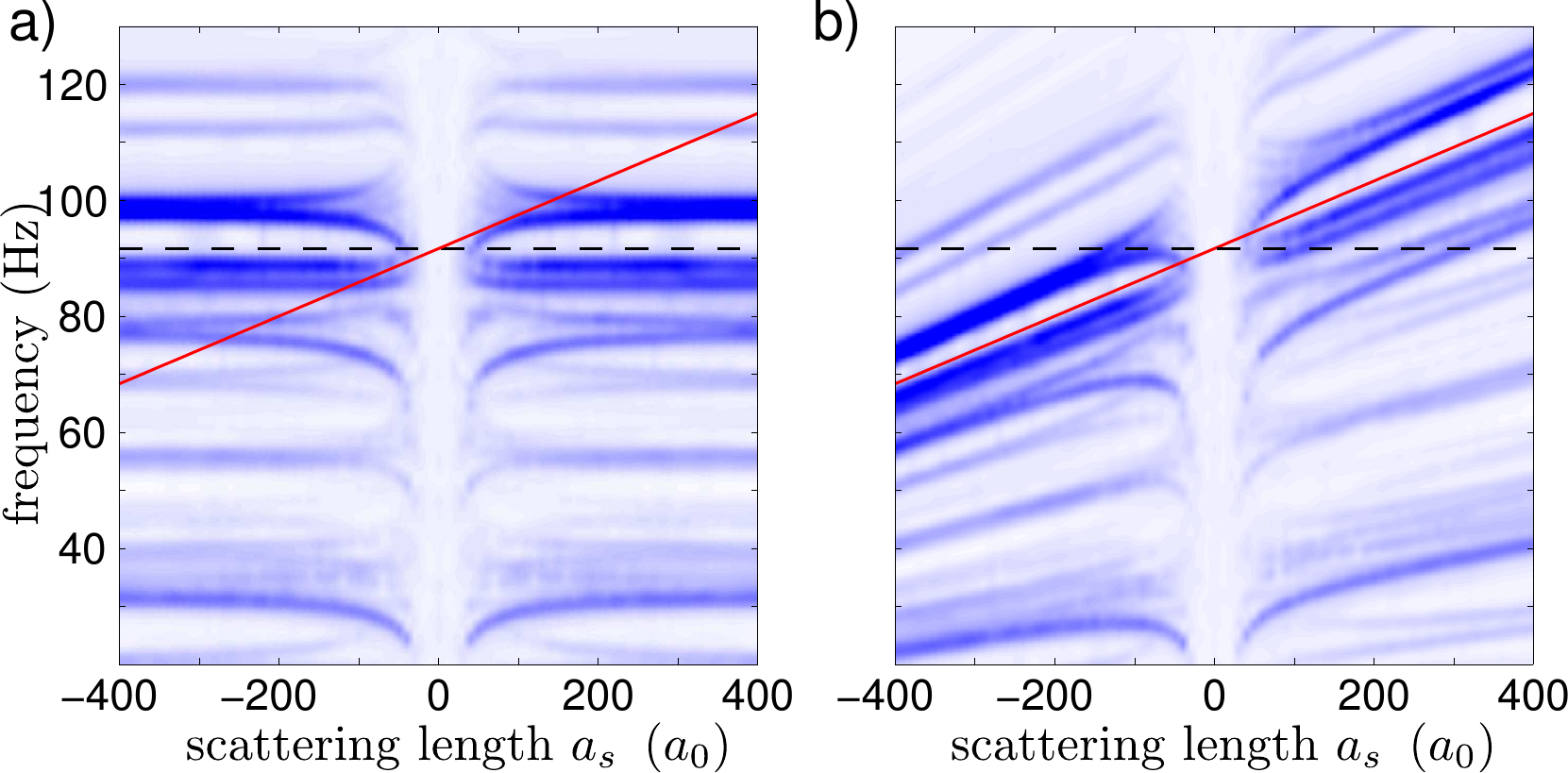}
\caption{Comparison of mode spectra as a function of the scattering length between (a) the standard Hubbard model and (b) the generalized Hubbard model with density-induced tunneling. Results are shown for a chain of 8 sites at occupation $n=1$ and lattice depth $V_z=10\, E_R$. The dashed lines depict the Hubbard model prediction $f_0=4J/h$, whereas the solid lines are modified by density-induced tunneling $f_0=4 (J + \Xone)/h$. \label{fig_standard}}
\end{figure}

In the experiment at resonant tilt $E=U$ the on-site interaction $U$ for a tunneling event in a Mott insulator is exactly compensated for by the tilt energy $E$. Consequently, no dependence of the dynamics on the scattering length is expected within the standard Hubbard model (dashed lines in \Fig{fig_standard}). This is confirmed by our numerical simulations without density-induced tunneling, shown in \Fig{fig_standard}(a). In contrast the generalized Hubbard model with density induced tunneling clearly shows the expected dependence of the resonant frequency on the scattering length (solid lines).

\subsection{Fourier analysis of doublon dynamics at occupation $n=2$}

\begin{figure}[t]
\centering
 \includegraphics[width=\linewidth]{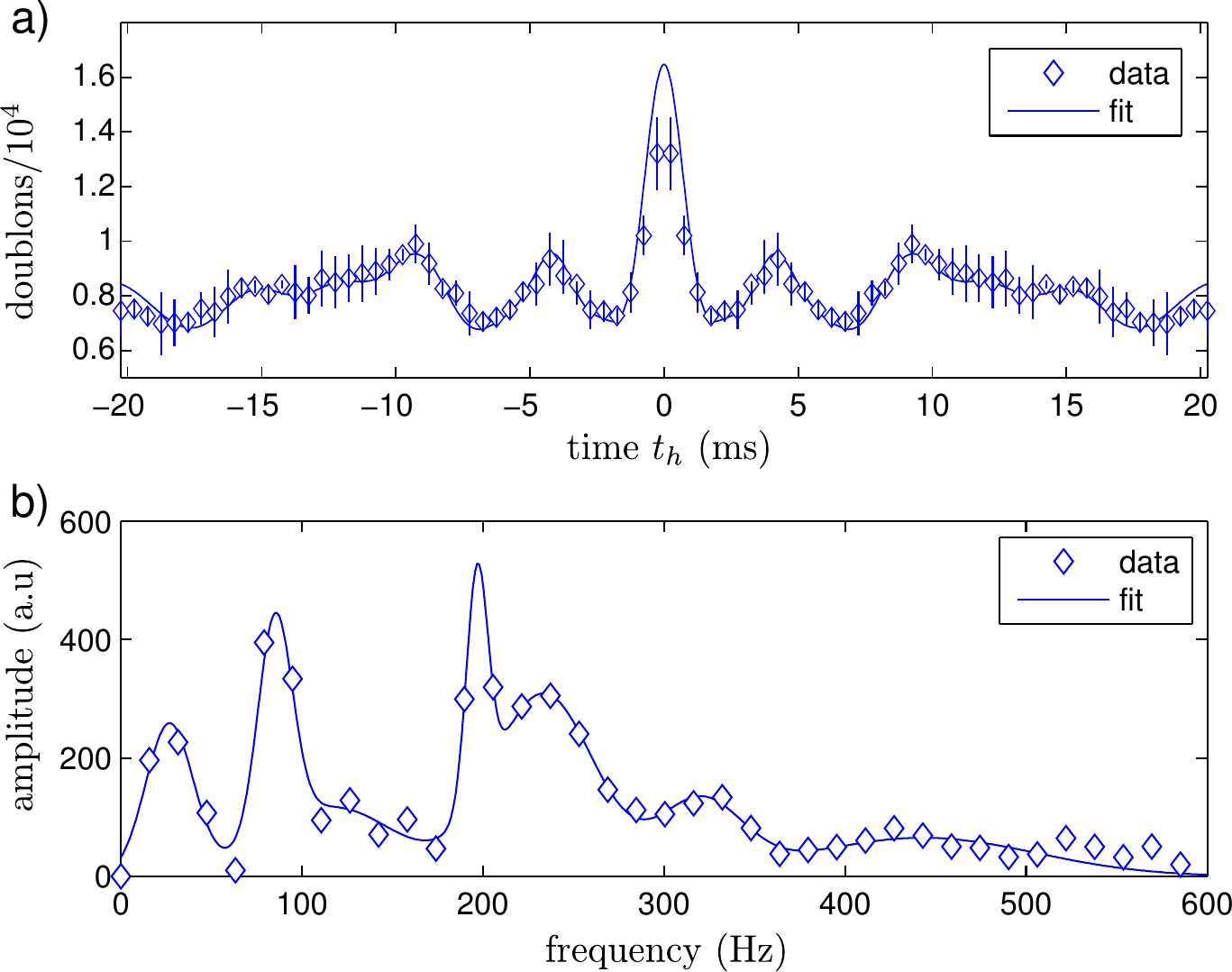}
\caption{(a) Mirrored time trace of the doublon dynamics at $V_z=10\, E_R$ and $a_s = 100\, a_0$ for $n=2$. The solid line is the Fourier transform of the fit in frequency space shown in (b). (b) Frequency spectrum calculated via Fourier transformation of the data shown in (a). The solid line is a fit using multiple Gaussians. \label{fig_fourierAnalysis}}
\end{figure}

The dynamics of the $n=2$ Mott insulator involves several distinct frequencies that we identify via a Fourier analysis. The experimental data at the resonant tilt $E=U$, $V_z=10\, E_R$ and $a_s=100\, a_0$ is shown in \Fig{fig_fourierAnalysis}(a). We enhance the resolution by mirroring the data at $t_h=0$. We associate the first data point with a holding time of $\Delta t_h=1\,$ms allowing for equidistant data. The consecutive Fourier transformation is shown in \Fig{fig_fourierAnalysis}(b) together with a fit of multiple (in this case seven) Gaussians. This procedure allows us to extract the resonant frequencies from the doublon dynamics, which can be compared with the theoretical predictions (see \Fig{fig_triplonDynamics} of the main text). We attribute the lowest frequency resonance to losses. Especially broad resonances, such as the third and last one in \Fig{fig_fourierAnalysis}(b), probably stem from several resonances that cannot be resolved. From the fit in the Fourier space we reconstruct the time trace (solid line in \Fig{fig_fourierAnalysis}(a)) and compare it with the original data, which gives us an estimate of the quality of the fit.

\begin{figure}[t]
\centering
 \includegraphics[width=\linewidth]{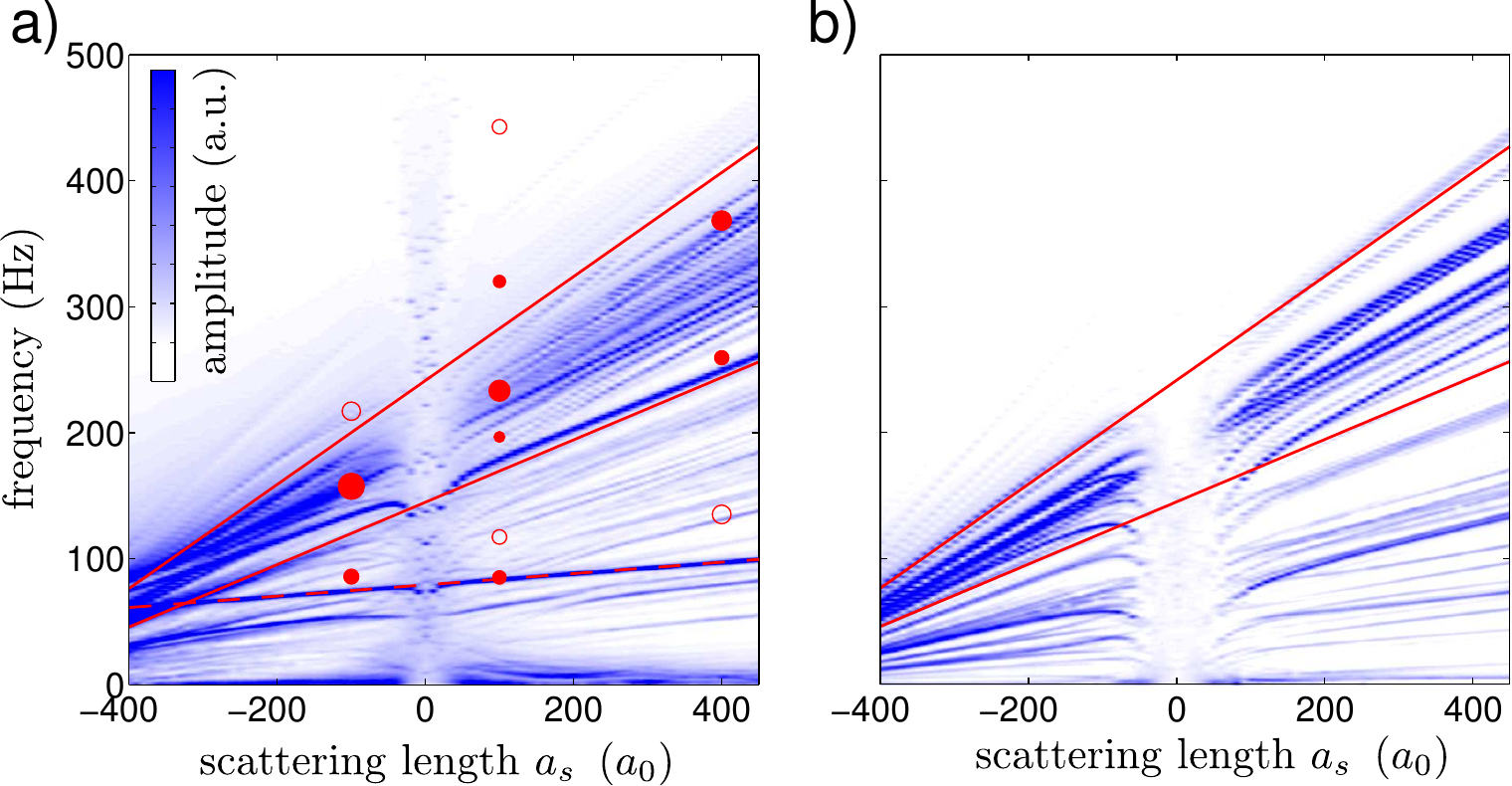}
\caption{Frequency spectra of the $n=2$ Mott insulator as a function of the scattering length for $V_z=9.6\, E_R$ and $E=U$. (a) Statistical average of chains $|2,...,2,0\rangle$. The solid lines mark the range $\nu (J+3\Xone)/h$ with $3\sqrt{3} < \nu < 5\sqrt{3}$. The dashed line in (a) shows the resonance at $2\sqrt{2} (J+\Xone)/h$. The markers are the resonances extracted from the experimental data (see \Fig{fig_triplonDynamics}(c) in the main text). The spectra are obtained from the first 2.5 s of the time evolution with a Gaussian broadening of $w=1$ Hz. (b) Results for a defect free chain with 8 sites (compare \Fig{fig_triplonDynamics}(c) in the main text). \label{fig_defects}}
\end{figure}

\subsection{Role of defects for $n=2$}
After the applied cleaning procedure, we estimate an amount of $p=20\%$ randomly distributed empty defect sites in the initial $n=2$ Mott insulator shell. Consequently, the one-dimensional lattice is divided into chains of the form $|0,2,...,2,0\rangle$, i.e., doubly occupied sites bounded by hole defects, where the tilt energy increases from left to right. Within the doubly occupied domain the process $|2, 2 \rangle \leftrightarrow |3, 1 \rangle$ is dominant and on the right side of the chain the process $|2,0\rangle \leftrightarrow |1,1\rangle$ takes place. Tunneling processes to the leftmost site are generally off resonant, since they are associated with a gain in tilt energy that cannot be compensated by an increase in interaction energy. Therefore, the leftmost hole site does not contribute to the dynamics within this subsystem. As a consequence, the hole site on the left side separates the one-dimensional lattice into chains of the form $|2, ..., 2,0 \rangle$, which can be treated separately. The only exception are chains with only one occupied site, where the transition $|... 2,0,2,0 \rangle \rightarrow |... 1,1,1,1\rangle$ can occur via two resonant processes.

Neglecting this coupling and higher-order processes, we simulate the effect of hole defects by statistically summing up separated chains $|2,...,2,0\rangle$ of a total length of $N$ sites, where we assume a relative probability $P_N=p\, (1-p)^{N-1}$. The dynamic behavior of chains with length $N>9$ is approximated by that of the $N=9$ chain, which is the upper limit for our numerical study.
Figure \ref{fig_defects} shows a comparison between the dynamics of the statistically averaged chains and of a defect free chain with $N=8$ sites as discussed in the main text. In both cases we observe a broad band of frequencies at $\nu (J+3\Xone)/h$ with $\nu$ in the range $3\sqrt{3} < \nu < 5\sqrt{3}$  (red solid lines) and a number of weaker resonances with smaller values of $\nu$. The factor $3$ in front of the density-induced tunneling term $\Xone$ allows to associate the resonances with the main process $|2,2\rangle \leftrightarrow |1,3\rangle$. In addition, a frequency mode with  $2\sqrt{2}(J+\Xone)$ appears for the defect averaged spectrum (dashed red line), which stems from the $|2,0\rangle \leftrightarrow |1,1\rangle$ process in $N=2$ chains. In the experiment, it is expected to be suppressed and shifted to higher frequencies for increasing $|a_s|$ due to multiorbital effects.

We conclude that the dynamics of the $n=2$ Mott insulator is not affected qualitatively by defects and thus we restrict our discussion in the main text to defect free simulations.

\end{document}